\documentclass[review,sort&compress]{elsarticle}

\usepackage{fancyhdr}
\usepackage[dvipsnames]{xcolor}
\usepackage{lineno}
\usepackage{booktabs}
\usepackage[colorlinks = true,linkcolor = blue]{hyperref}
\usepackage{amsmath}

\usepackage{float}
\usepackage{amssymb}
\usepackage{subcaption}
\usepackage{graphicx}
\graphicspath{{Images/}{./Images/}} 

\usepackage{caption}
\usepackage{subcaption}

\usepackage[strict]{changepage}

\usepackage{framed}

\definecolor{formalshade}{rgb}{0.95,0.95,1}

\newenvironment{formal}{%
	\MakeFramed{\advance\hsize-\width\FrameRestore}%
	\noindent\hspace{-4.55pt}
	\begin{adjustwidth}{}{7pt}%
		\vspace{2pt}\vspace{2pt}%
	}
	{%
		\vspace{2pt}\end{adjustwidth}\endMakeFramed%
}

\journal{Elsevier}

\begin{document}

	\def\bibsection{\section*{References}}
	\renewcommand{\labelitemii}{$-$}
	
	\begin{frontmatter}
		
		
		\title{A graph based workflow for extracting grain-scale toughness from meso-scale experiments}

		\author[TEC]{Stylianos Tsopanidis\corref{cor}}
		\author[TEC]{Shmuel Osovski}
		\cortext[cor]{Corresponding author. Tel. +972-512820572; E-mail address: tsopanidis@campus.technion.ac.il}

		\address [TEC]{Faculty of Mechanical Engineering, Technion - Israel Institute of Technology, Haifa, Israel}

		\begin{abstract}
		 
		We introduce a novel machine learning computational framework that aims to compute the material toughness, after subjected to a short training process on a limited meso-scale experimental dataset. The three part computational framework relies on the ability of a graph neural network to perform high accuracy predictions of the micro-scale material toughness, utilizing a limited size dataset that can be obtained from meso-scale fracture experiments. We analyze the functionality of the different components of the framework, but the focus is on the capabilities of the neural network. The minimum size of the dataset required for the network training is investigated. The results demonstrate the high efficiency of the algorithm in predicting the crack growth resistance in micro-scale level, using a crack path trajectory limited to 200-300 grains for the network training. The merit of the proposed framework arises from the capacity to enhance its performance in different material systems with a limited additional training on data obtained from experiments that do not require complex or cumbersome measurements. The main objective is the development of an efficient computational tool that enables the study of a wide range of material microstructure properties and the investigation of their influence on the material toughness.

		\end{abstract}


	\end{frontmatter}
	
	\section{Introduction}
	Interface controlled fracture is an important phenomenon present in many structural alloys such as Al-Li \citep{Messner2014, lynch1991fracture}, $\beta$-Titanium alloys \citep{osovski2015}, multi-phase steels \citep{Liu2020g} and ceramics \citep{feng2018influence,zhou2004stochastic,auger2016crack}. Similarly, when exposed to aggressive chemical agents (e.g. sulfur, liquid metals, hydrogen etc.), both metallic and ceramics materials will exhibit an increasingly growing susceptibility for grain boundary fracture \citep{watanabe2004toughening,kobayashi2014situ,sun2019grain,miura2015micro}. Moreover, many composite materials (polymer or metallic based) are prone to fail along interfaces defined by their internal structure \citep{mueller2015quantification, hilditch2007effect,yuan2020influence,abdullah2015interfacial,jia20193d}. As such, being able to correlate a material's microstructure with its crack growth resistance in a quantitative manner is an open challenge with significant implications on the field of materials design. 
	
	Despite the rapid advancement in experimental techniques, which now days span multiple time and length scales, high-throughput measurements, probing the local toughness of a material's interfaces, are scarce. In situ observations are mostly limited to either extremely small scales \citep{miura2015micro,feng2018influence,ast2019review,Sangid2020}, or extrapolating between local and global measurements via finite elements simulations \citep{alabort2018grain}. 
	On the other-hand, new simulation techniques, the availability of high performance computers, reduced order models and data-driven approaches are allowing  to numerically probe the effects of various microstructure descriptors, including the grain size and grains orientation distribution, thus paving the way to a computationally informed material design \citep{eghtesad2020high,bachurin2018influence,roy2020,molkeri2020,osovski2015,Roy2021,li2021transgranular,shterenlikht2018modelling,MontesdeOcaZapiain2021,Mangal2018,Hunter2018,Panda2020a,Gomberg2017,Liu2020e,Pierson2019,Gupta2019,jodlbauer2020parallel,heister2020pfm}. 
	
\begin{center}
		\textit{The lack of sufficient experimental data and the difficulty in obtaining it at the desired scales stands at the basis of the presented work. }
\end{center}

	In a recent work,  \citet{osovski2019} and \citet{molkeri2020} utilized a graph based model for crack growth along grain boundaries. The model, introduced in \citet{osovski2015} to explore the role of the grain boundaries network on fracture in meta-stable $\beta$ Titanium alloys, is based on treating the crack growth process as a collection of distinct unit events where a crack grows from one grain boundaries junction to another. The aforementioned approach, while highly efficient from a computational perspective, is very limited due to the difficulties in obtaining experimental data as to the relation between the grain boundaries character, geometry and the resulting local crack growth resistance.

\begin{center}
		\textit{The work presented here is aimed at devising a method for extracting \textbf{micro-scale} toughness information from \textbf{meso-scale experiments} while capitalizing on proven methodologies from the field of data-science. }
\end{center}

	 In the sections to follow, a three components modular computational framework is described and analyzed as to its ability to estimate the crack growth resistance of local interface segments from global $J-\Delta a$ experimental curves. While we rely heavily on the methodology used in \citet{osovski2015}, \citet{osovski2019} and \citet{molkeri2020}, this assumption is made solely for keeping the computational costs low, and the same procedures can be followed while utilizing a variety of other computational or theoretical methods (e.g \citep{microfract,molkeri2020,graph2021,Roy2021} ).

\begin{figure}[!h]
		\centering
		\includegraphics[width = 0.8\linewidth]{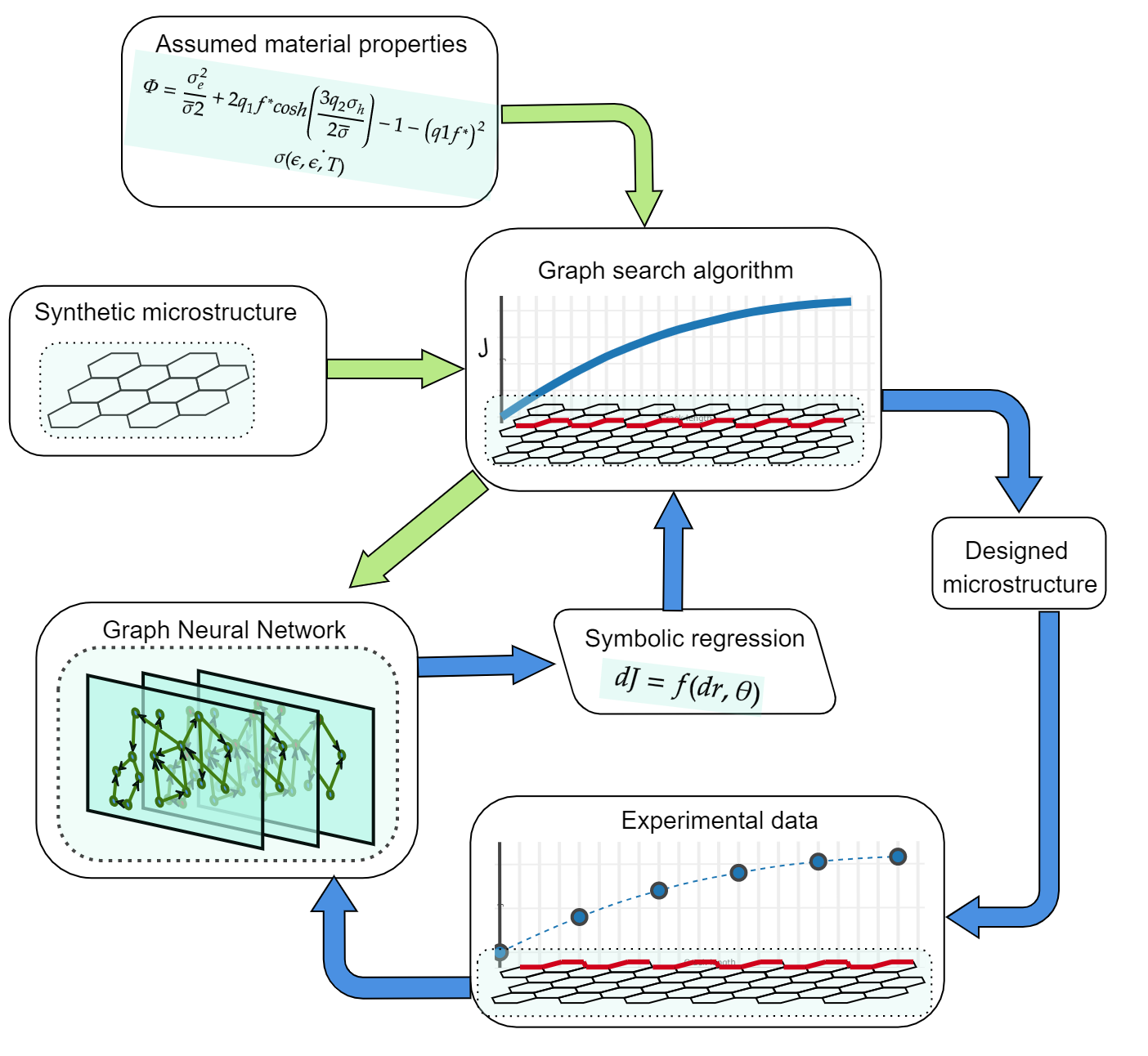}
		\caption{Flowchart of the envisioned computational framework integration with experimental data. First, a GNN is trained on synthetic data using the graph search algorithm in \citet{osovski2015} and learns to convert the macroscopic $J-\Delta a$ curve into an analytical expression. Next, the GNN is retrained using experimentally obtained data leading to a new analytical expression while requiring significantly smaller data-set for re-training. Finally, the newly obtained analytical expression can be used for optimizing the grain size distribution. }
		\label{fig:flowchart}
	\end{figure}

	 The proposed method ,presented schematically in Fig. \ref{fig:flowchart}, is based on the following assumptions:

\textit{	 \textbf{Assumption 1}: 
	The material to be studied fails by crack growth along grain boundaries or other forms of interfaces and thus has a pre-defined network of routes on which the fracture process takes place. 
}
	 
\textit{	 \textbf{Assumption 2}: 
		It is possible to obtain experimental data, in the form of $J-\Delta a$ with a spatial resolution of the order of several interface segments.
}

	\textit{ \textbf{Assumption 3}: 
	 The experimentalist has a general idea as to the physical factors governing the local behavior of the crack propagation  (e.g., geometrical factors, grain boundary character, etc.).}

	We note that while Assumption 3 may seems to pose a strong limitation for the proposed methodology, it is actually quite the opposite.  The exact functional form itself is of less importance as long as the input variables have been identified, and it is possible to formulate (either analytically or computationally) a relation between the crack path selection and the local toughness.

	Given that the aforementioned assumptions are valid, the proposed framework can extract the micro-scale toughness, in the form of a function tying the assumed input variables with the crack growth resistance, from experimentally obtain $J-\Delta a$ curves at the meso-scale and optical or electron microscopy of the fractured specimen. 

    To this end, our framework combines three computational parts:
	\begin{itemize}
    \item  \textbf{Database Generation}: a synthetic database is created using the assumed functional relation and microstructures of relevance to the actual material.
	\item \textbf{Model Training}: a machine learning algorithm, based on a graph neural network architecture, is trained on predicting the crack growth resistance curve for an input crack path trajectory (i.e. sampling of the input physical variables). The training phase is performed in two stages:
		\begin{enumerate}
			\item The training is performed on an extensive dataset of crack path segments (a segment is the equivalence of spacing ($\Delta a$) between experimentally collected $J$ sampling points).
			\item An additional training cycle is conducted on the already trained weights using as input a sequence of crack path segments obtained experimentally. 
		\end{enumerate}
	 \begin{formal}
	 	Although the training is based on predicting the value of $J$ required to transverse a segment(typically 5-15 grains in the examples to follow), the network is capable of predicting the local crack growth resistance increments ($dJ$) with high accuracy, as is shown in Section 3. 
	\end{formal}
	\item \textbf{Symbolic Regression}: The fully trained network is used to generate a dataset relating the \textbf{local} grain boundary properties and associated value of $dJ$ (the energy required for a crack to grow along the grain boundary). A symbolic regression code is then used to compute an analytical mathematical expression relating the properties chosen as input variables (in our example the length, $dr$ of a grain boundary and its angle ($\theta$) with respect to the loading direction) with $dJ$.  The obtained expression is thus a representation of the experimental data points and can be used for further computational schemes or optimization tools for microstructure design.
\end{itemize}

It is important to note that while the first stage is critical for the proposed framework, its only purpose is to allow for transfer learning in the second stage, such that minimum training will be required to produce accurate predictions. In other words, the sole purpose of the first stage is to avoid the necessity of producing an extremely large amount of experimental data \citep{verma2019learning}.

	The rest of the manuscript is organized as follows: 
	We start by presenting the training and validation dataset along with the algorithm used for his generation. Subsequently, the theoretical formulation of the graph neural network algorithm is explained and the underlying mathematical operations that enable its functionality are elucidated. 	Next, the predictive capabilities of the algorithm on an unseen dataset, are presented and discussed in light of the distribution of parameters in the training dataset. Evidently, the functionality of the proposed framework relies on the fact that the prediction performance of the GNN can be enhanced for a new material system with a limited additional training. Thus, we explore  what is the minimum amount of crack path data needed to enable the neural network to perform reasonably accurate predictions on a new material system, and discuss the algorithms accuracy in relation to the variations introduced to the data generation algorithm and the size of the new dataset. 
	 
	\section{Methods}\label{Methods}
	
	\subsection{Dataset}
	 The first part of the development of a computational framework based on a machine learning algorithm involves the creation of an extensive dataset, whose main part will be used to train the neural network and a much smaller fraction is kept for the validation of the network's accuracy. As a means to demonstrate the proposed framework and test its capabilities and limitations, we have chosen to use the methodology in \citet{osovski2019} as our physical assumption regarding the local crack growth behavior. 
	 
	 Following \citet{osovski2019}, a centroidal Voronoi tessellation scheme is implemented in order to generate a dataset of microstructures. This dataset consists of microstructures with randomly oriented grains in a two-dimensional plane. The microstructures are subsequently converted into directed graphs in \textit{Euclidean} space, where the graph nodes represent the grain boundaries triple points, and the graph edges represent the sides of each grain. The node features on this graph are assigned to the spatial coordinates of each junction, while the edge features are set to the length of the grains' sides and the angles between each side and the horizontal axis of the coordination system. An additional auxiliary graph in $J$-resistance space, which stores the information of the crack growth resistance values ($dJ$) between any pair of nodes in the directed \textit{Euclidean} graph, is also created. The $J$-resistance between two consecutive nodes depends on the length (\textit{dr}) of the edge that connects them and the edge angle (\text{$\theta$}) with the horizontal axis; a relation $ J = f(dr,\theta)$ is defined according to discrete unit event Finite Element calculations. As described in \citet{osovski2015} and \citet{osovski2019}, unit event finite element simulations, where the angle ($\theta$) between two grain boundary edges and the initial edge length ($dr$) are varied, is implemented in order to interpolate a function that relates $dJ$ with the topological characteristics of the grain boundaries and their junctions. The analytic mathematical expression that relates these variables, after being calibrated with finite element modeling, is defined as follows:
	 
	 \begin{equation}
	 dJ= \left\{
	 	\begin{array}{ll}
	 	(A_1|\theta|^{4.397} + A_2)\cdot|dr\cdot cos(\theta)| + B_1\theta - B_2, \; if \; \theta > 0.96 \; rad \\
	 	\\
	 	(A_1|\theta|^{4.397} + A_2)\cdot|dr\cdot cos(\theta)|, \: otherwise
	 	\end{array} 
	 	\right.
	 \label{eq:dj}
	 \end{equation} 
	  
	where the coefficients are defined as: $A_1 = 5.376,\; A_2 = 3.6, \; B_1 = 0.010313\; \text{and} \; B_2 = 0.0098$.
	
	Combining the information provided by these two graphs allows the computation of the path that the crack follows starting from any predefined initial point of a microstructure. In each grain boundary junction of the microstructure (or else each node of the graph) the crack path propagates towards the direction of the minimal J-resistance value. Thus, for each microstructure generated by randomly positioned and oriented set of grains, a crack path that minimizes the crack growth resistance, starting from the nodal point closest to the middle of the left side of the microstructure, is computed. For each crack path, which is defined by the \textit{Euclidean} coordinates of the grain boundary junctions that belong to this path trajectory, the $dJ$ values that correspond to the crack growth resistance between two consecutive junctions are computed. The summation of all these energy increments ($dJ$) computes the global $J$ value for the entire path.

	Considering the experimental methodology \citep{ASTMJ} used for determining the crack growth resistance curve ($J-\Delta a$), the final dataset produced using the graph search algorithm contains 40000 crack path segments, each one containing $8$ grain boundary edges, and the corresponding sequence of $dJ$ values, which stores the $dJ$ increments for each grain in the each crack segment. These crack segments simulate the crack propagation during the loading intervals in the experimental methodology and discretize the entire crack path into a sequence of consecutive crack segments -- each segment corresponding to crack propagation through 8 grain boundary edges. During the graph neural network training, the training input is the crack path segment, defined by the junction coordinates, and the training target is the $J$ value for this specific crack path segment, which corresponds to the sum over the 8 local $dJ$ increments  (see Eq.~\eqref{eq:global_J}). The local energy values ($dJ$) are only used during the evaluation of the network's efficiency analysis, as it will be explained later.
	
	\begin{equation}
		J = \sum_{i=1}^{8}dJ^i
		\label{eq:global_J}
	\end{equation}

	\subsection{Graph Neural Network}
	
	\subsubsection{Theoretical overview}
	
	Considering that the input data is formulated in the graph domain, the neural network architecture proposed for processing and consequently performing predictions on this dataset should be invariant to node or edge permutations of the input graph structure. The idea of a Graph Neural Network (GNN) was first introduced by \citet{gnn1} and further developed by \citet{gnn2}. These GNN algorithms were further optimized \citep{gnn3,gnn4,gnn5,gnn6,gnn7,gnn8} to reduce the computational cost and increase their training efficiency and representation capability. Finally, \citet{battaglia} presented a generalized version of a Graph Network, integrating different GNN models among which the Message Passing Neural Network \citep{gnn7} and Non-local Neural Network \citep{gnn8}. The proposed model, which is implemented to produce the results presented here, exhibits high efficiency in capturing the relational interconnections between the different elements (nodes, edges) of the graph structures and embedding the input graph features in high-dimensional latent space, which enable high accuracy predictions.
	
	Following the framework introduced in \citet{battaglia}, a graph $G = (u, V, E) $ is defined by:
	\begin{itemize}
		\item a global features vector $ u $, which embeds the entire graph representation into a single feature in the latent space,
		\item a set of node features $ V = \{v_{i}\}_{i=1:N^{v}}$, which in the crack path dataset are assigned to the spatial coordinates of each grain boundary junction that is part of the crack,
		\item a set of edge features $ E = \{(e_{k}, r_{k}, s_{k})\}_{k=1:N^{e}}$, where $ e_{k} $ represents the edge features of the edge $k$ and in the case of the crack path dataset these features are the edge length ($dr$) and the edge angle ($\theta$)  -- for each edge that belongs to the path. Additionally, $s_{k}$ and $r_{k}$ denote the sender and receiver nodes of the $k^{th}$ edge. 
	\end{itemize}  
	
	The GNN uses three Multi-layer Preceptron (MLP) networks that act as update functions for the edge, node and global features of the input graph and convert them into latent representations. The operation of the network, defined in detail in \citet{battaglia}, can be summarized as follows:
	\begin{enumerate}
		\item the edge update MLP ($\phi^{e}$) operates on a vector that concatenates the features of each edge ($e_k$), the features of the sender  and receiver nodes of the edge ($v_{r_k}, v_{s_k}$) and the global feature ($u$). This operation can be mathematically formulated by:
		\begin{equation}
			e'_k = \phi^e ([e_k, v_{r_k}, v_{s_k}, u])
		\end{equation}
		The edge update function $\phi^{e}$ acts on the features that correspond to each edge, using the same set of weights, and the result ($ e'_k$) of this operation is the updated feature vector of the edge. The final results of this operation is a set of updated edge features: $E' = \{(e'_{k}, r_{k}, s_{k})\}_{k=1:N^{e}}$

		\item After updating the features of every edge in the input graph, all the features of the edges that share the same receiver node ($r_k=i$) are summed to produce a new set of features $\{\bar{e}'_i\}_{i=1:N^v}$. Each element of the set integrates the edge information of all the edges, connected to each node, to a single feature and it will subsequently be used for updating the node features at the next stage of the computational framework. 
		
		\item The node update MLP ($\phi^{v}$) is applied to the feature vector that is constructed from the concatenation of the summed edge features per node ($\bar{e}'_i$), the node features ($v_i$) and the global feature ($u$). The elements of the updated set of node features ($V' = \{v'_i\}_{i=1:N^v}$) are computed by:
		\begin{equation}
			v'_i = \phi^v ([\bar{e}'_i, v_i, u])
		\end{equation}
		
		\item The updated set of edge features $E'$ is summed to produce a single global edge feature vector ($\bar{e}'$). Similarly, the updated set of node features ($V'$) is summed to a single global node feature vector ($\bar{v}'$). This step of the computational framework enables the integration of the information encapsulated into the updated edge and node features into two global features.
		
		\item The global features update MLP ($\phi^u$) operates on the concatenation of the aggregated edge features attribute ($\bar{e}'$), the aggregated node features attribute ($\bar{v}'$) and the global feature, resulting into the updated global feature:
		
		\begin{equation}
			u' = \phi^u([\bar{e}', \bar{v}', u])
		\end{equation}     
	\end{enumerate}
	
	\subsubsection{Network Architecture and Training parameters}
	
	The design of the GNN architecture accounts for two important factors:
	\begin{itemize}
		\item the neural network should have sufficient number of layers and each layer sufficient depth in latent space that enables the encoding of the node and edge features and captures their inter-connectivity,
		\item the neural network should have the simplest possible structure, in terms of number of layers and layer size, in order to allow the training of the network with the smallest possible training dataset size.
	\end{itemize}

	Apparently, these two competitive factors should be considered and the neural network design should be optimized to generate an algorithm that is capable of achieving high accuracy predictions with minimum training. This will allow the neural network to be extended to perform accurate predictions on a dataset that deviates from the initial training dataset with minimal additional training.
	
	Taking  under consideration these factors the general structure of the GNN is composed by three graph network building blocks:
	
	\begin{enumerate}
		\item the \textbf{encoder graph network} is used to encode the node and edge features into high-dimensional representations in latent space. The implementation of a node (or edge) MLP, containing a single fully-connected layer with 100 neurons, encodes the node (or edge) features into a 100-dimensional representation in the latent space. Thus, the final product of this building block is two 100-dimensional arrays expressing the features of each node and each edge of the input graph in the latent space representation.
		\item the \textbf{processing graph network} is the most important part of the GNN and its role is to capture the inter-connections between the node and edge features as they are encoded in the latent space. It consists of 3 MLPs (\textit{global}, \textit{edge} and \textit{node} update features operators) with a single fully-connected layer of 100 neurons and 3 corresponding aggregators that perform feature summations, following the computational framework described in the previous section. 
		\item the \textbf{decoder graph network} is the final part of the GNN architecture and its role is to decode the feature representations, as they are output from the previous processing block. The final product of this part is an 1-dimensional array with the predicted $dJ$ values for each edge of the input crack path graph.
	\end{enumerate}  
	
	The weights of all the MLP layers in the GNN architecture are trained using a mean square error loss function between the predicted ($J^{pred}$) and the ground truth ($J^{gt}$) values of each input crack path segment:
	
	\begin{equation}
		Loss = \frac{1}{N_{segments}} \sum_{i}^{N_{segments}} (J^{pred} - J^{gt})^2
		\label{eq:loss}
	\end{equation}
	
	where $N_{segments}$ is the number of input crack segments of the crack path. The $J^{pred}$ and $J^{gt}$ are the predicted and ground truth crack growth resistance values for each crack path segment, which consists of 8 grain boundary edges, and are computed by the sum of the $dJ$ increments at each grain boundary edge that belongs to the specific segment:
	
	\begin{equation*}
		J = \sum_{k=1}^8 dJ^k
	\end{equation*}

   	It is worth-noticing that the GNN algorithm is capable of predicting the local $dJ$ increments at each crack path edge, while it is only trained on the $J$ value of the entire crack path segment. This training strategy is followed to be inline with readily accessible  experimental techniques, as the final objective of this work is the development of a framework which enables predictions on experimental data, after a short training process on a sub-set of this experimental dataset.

	\section{Results and discussion}
	
	\subsection{Training and prediction accuracy on the initial dataset}
	
\begin{figure}[!h]
		\centering
		\includegraphics[width = 0.95\linewidth]{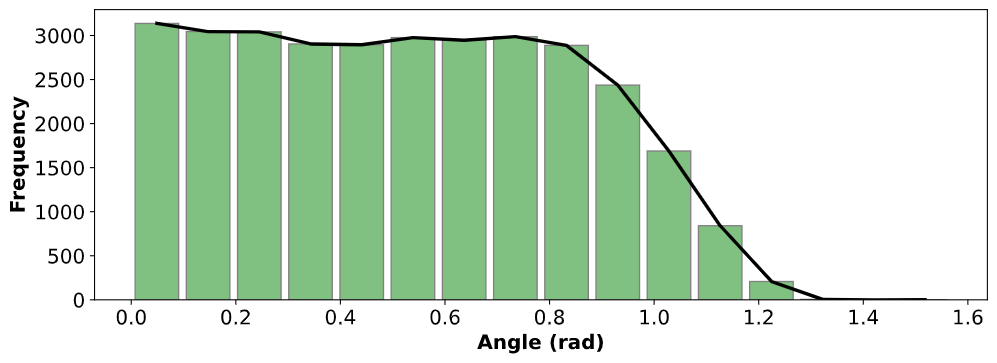}
		\caption{The angle between the grain boundary sides and the horizontal axis distribution in the crack path dataset.}
		\label{fig:angle_distrib_init}
	\end{figure}

	Initially, the graph neural network is trained on an extensive dataset; 40,000 input crack segments, with 8 grains in each segment, result in 320,000 grains belonging to the crack path. The dataset is split into training and validation sets, with 90\% of the data assigned to training dataset and the rest to the validation dataset. The distribution of the grain boundary angles ($\theta$) belonging to the input crack path is presented in the histogram plot in Fig. \ref{fig:angle_distrib_init}. As expected, the probability of finding angles larger than 0.96 rad is much lower than the probability for smaller angles, since the graph search algorithm, implementing Eq.~\eqref{eq:dj}, favorites smaller angles when searching for the crack path in the material microstructure that minimizes the overall $J$-resistance value. Furthermore, angles larger than 1.2 rad are not present in the crack path, because the energy increments that correspond to this angle range are very large.  
	
	The weights of the network are trained using Adam stochastic gradient descent optimization method \citep{adam} with learning rate gradually reduced during training from $5\cdot 10^{-4}$ to $10^{-5}$, using the default moment parameters ($\beta_1 = 0.9$ and $\beta_2 = 0.999$). The training is using as inputs the crack path trajectories of the training dataset and the network weights are updated in order to minimize the loss function, computed between the predictions of the $J$ values of the input crack path segment and their ground truth values, provided by the targets of the dataset (see Eq.~\eqref{eq:loss}). During the validation phase at the end of each epoch the performance of the algorithm on unseen data is evaluated in order to prevent \textit{over-fitting}. 
	
	The network is trained for 220 epochs on a personal computer, equipped with an NVIDIA\textsuperscript{\textregistered} GeForce\textsuperscript{\textregistered} RTX 2080 Ti Graphics Processing Unit (GPU) and the mean training time for each epoch is 320 $sec$. The prediction accuracy of the algorithm is evaluated on a test dataset, consisting of 2,000 crack path segments, with 8 grains in each segment, leading to a total of 16,000 grains in the test dataset. 
	
	The mean accuracy on predicting the $J$ values for the input crack path segments is computed to 97.29\%.
	
	Additionally, the neural network is capable of predicting the local increments ($dJ$) from one grain boundary junction to the next, even though it is not trained on the prediction of those values, but rather on the prediction of the total crack growth resistance of a segment. The mean accuracy in the prediction of the local $dJ$ increments is evaluated to be 87.51\%, which is very accurate considering that the network was not given this information in the training process.

\begin{figure}[!h]
		\centering
		\begin{tabular}{c c}
			\includegraphics[width = 0.48\linewidth]{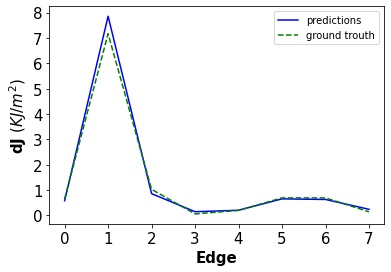} &
			\includegraphics[width = 0.48\linewidth]{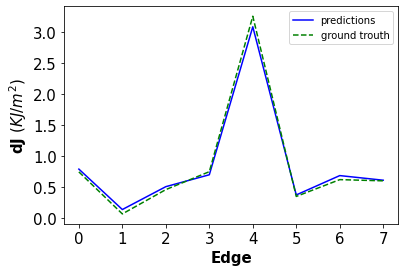} \\
			\includegraphics[width = 0.48\linewidth]{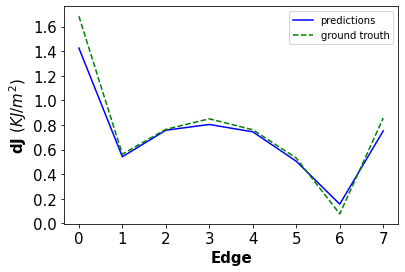} &
			\includegraphics[width = 0.48\linewidth]{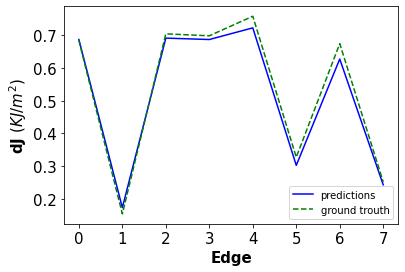} 
		\end{tabular}
		\caption{Ground truth and predictions of the local $dJ$ increments between a grain boundary junction and the consecutive junction in the crack path.}
		\label{fig:j_init_plots}
	\end{figure}

	Fig. \ref{fig:j_init_plots} presents the prediction and ground truth values for the J-resistance increments for some characteristic cases of the test crack path dataset. The pronounced peaks on some of the plots of the Fig. \ref{fig:j_init_plots} correspond to the grain boundary edges for which $\theta > 0.96$ rad. 
	
\begin{figure}[!h]
		\centering
		\includegraphics[width = 0.95\linewidth]{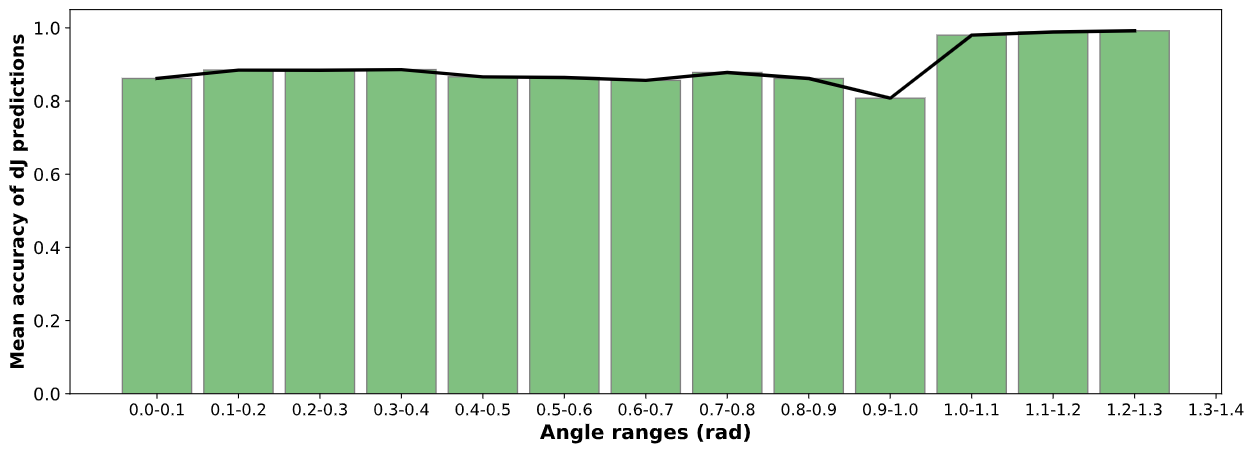}
		\caption{The prediction accuracy for the local $dJ$ increments with respect to the grain boundary angle $\theta$.}
		\label{fig:angle_acc_init}
	\end{figure}

	Additionally, the dependence of this accuracy on the angle $\theta$ was investigated. The mean accuracy in $dJ$ increments for different angle ranges is computed and presented in the bar plot in Fig. \ref{fig:angle_acc_init}. The plot shows that the mean accuracy on the predictions of the local $dJ$ increments remains high in all angle range groups, exhibiting almost perfect accuracy on the cases where$\theta>0.95$ rad, which is the threshold value in the branched Eq.~\eqref{eq:dj}. 
	
	The prediction accuracy of the network on the $J$ value for the entire input crack path segments is very high, since it is trained on an extensive training dataset. Moreover, the prediction accuracy for the local $dJ$ increments is also high, which is very important considering that the network is trained only on the prediction of the total $J$ value for the crack path segments, (the loss is based on the evaluation of $J$ value for the path segment, as it is defined in Eq.~\eqref{eq:loss}).

	\subsection{Performance on different datasets}
	
	The proposed computational framework is of use for the mechanics of materials community only if it can expand its performance in computing the crack growth resistance measures for data extracted from experimental studies. Thus, this section focuses on exploring the enhancement of the performance of the algorithm on synthetic data that are produced from variations of the Eq.~\eqref{eq:dj}. Altering the coefficients or even the structure of this equation results in different crack path trajectories and different J-resistance values, hence the dataset produced with this method can substitute the dataset obtained by experimental measurements. Our objective is to understand how extensive should be the additional training of the network in order to perform high accuracy predictions on a dataset exhibiting a different dependence of the local toughness on the input variables than the one it was trained on.

	The optimization process that is followed searches for the minimum crack path size, quantified by the number of crack path grains, that is needed for additional training of the network in order to enable predictions of the $J$ values for the crack path segments and their local increments with high accuracy. As before the training inputs are the segments of the crack path trajectory and the meso-scale $J$ values that correspond to these segments. 
	
	To this end, the network is subjected to additional training cycles, where the weights at each training are initialized with the coefficients that have been previously trained on the extensive dataset, as described in the previous section. Each of these training cycles uses as input a crack path with different sizes, which similarly to the previous is divided into several crack segments. After each training cycle is completed, the evaluation of the prediction performance of the network relays on the evaluation of its accuracy on the prediction of the $J$-resistance of the crack path segments, which is the operation that the network is trained on, and the accuracy on the prediction of the $dJ$ increments on each edge of the segments, even though the network is not trained on this task. Decreasing the size of the input crack path, or else the number of crack path grains, on each training session and evaluating the network's efficiency, enables the investigation of the minimum size for the input crack path, quantified by the number of grain edges that belong to it, that produces a trained network that performs high accuracy predictions.  
	 
	 Three different variations of the Eq.~\eqref{eq:dj} are studied:
	 
	 \begin{enumerate}
	 	\item \textit{Altering the angle dependence}: The exponent of the angle in Eq.~\eqref{eq:dj} is changed from 4.397 to 3 and the coefficient $A_1$ is changed to $A_1 = 0.00188$, instead of $A_1 = 5.376$ in the initial equation.
	 	\item \textit{Altering the structure of the equation}: The Eq.~\eqref{eq:dj} is converted from a branch equation to a normal equation by removing the term $B_1 \theta - B_2$.
	 	\item \textit{Combining the previous changes}: both previous changes are applied to the Eq.~\eqref{eq:dj}. More specifically, the equation that computes the crack growth resistance from a grain boundary junction to the next is defined as: 
	 	
	 	\begin{equation}
	 		dJ = (A'_1|\theta|^{3} + A_2)\cdot|dr\cdot cos(\theta)|
	 		\label{eq:dj_all}
	 	\end{equation} 
	 	
	 	where $A'_1 = 0.00188 $ and $A_2 = 3.6 $.
	 \end{enumerate}
	
	These changes not only affect the values of the $dJ$ increments of the crack path, as they are computed by the Eq.~\eqref{eq:dj}, but also alter the crack path trajectory itself, since this trajectory is determined by searching the path in the microstructure with the minimum crack growth resistance.
	
	Five different training datasets are produced by the graph search algorithm with 2000, 1000, 480, 320 and 240 grains, implementing each variation of the initial equation. Consequently, the network is subjected to five independent training sessions, using as starting training point the pre-trained weights on the extensive initial dataset. In parallel, a test dataset of a crack path with 4000 crack path segments (or else 32000 grains) is produced with each altered equation, in order to enable the evaluation of the performance of the algorithm after each training cycle is completed. 
	
	This iterative training and evaluation process is repeated for each alteration of the initial equation and the minimum dataset size that trains the network weights to performs high accuracy predictions on the test dataset is obtained. Table~\ref{table:min_dataset} presents the minimum training dataset sizes in terms of crack path grains for each alteration of the Eq.~\eqref{eq:dj} that ensure high accuracy predictions on the test datasets. Additionally, the extent of this additional training process is quantified by the number of training epochs. As the deviation of the training dataset with respect to the initial dataset increases, the neural network requires more extensive training in order to achieve high accuracy predictions.  Note that the small size of the training dataset reduces the training duration; on average each training epoch is completed in less than 7 sec, and even when the additional training involves 223 epochs, it is still a short process (approximately 26 min).    
	
	\begin{table}[!h]
		\centering
		\begin{tabular}[t]{ c |  c  c } 
			\toprule
			\toprule
            Equation variations       & 	\vtop{\hbox{\strut Minimum training dataset }\hbox{\strut (number of crack grains)}} & Training epochs  \\ \toprule 
		    Alter angle dependence     &  240  & 44 \\ \midrule
		    Alter equation's structure  & 320  & 112 \\ \midrule
		    Combine all the changes & 320 & 223 \\ \bottomrule  
			\bottomrule
		\end{tabular}
		\caption{Minimum size of the training dataset, quantified by the number of grains in the crack path, that is required to train the network to enable accurate predictions in test datasets produced by three different variations of the initial equation. The number of epochs that the additional training requires to produce an efficiently trained network is also presented.}
		\label{table:min_dataset} 
	\end{table}	
	
	As the crack growth resistance rule is subjected to more substantial changes, the network requires larger training dataset to be trained efficiently, although the increase on the dataset's size is not large (from 240 to 320 crack path grains).

	The evaluation of the network's accuracy on the test datasets is performed in two different scales:
	
	\begin{itemize}
		\item \textbf{Meso-scale toughness}: evaluated by computing the mean accuracy on the prediction of $J$-resistance of the test dataset's crack segments.
		\item \textbf{Micro-scale toughness}: evaluated by computing the mean accuracy on the local energy $dJ$ increments between two consecutive grain boundary junctions in the crack path.  
	\end{itemize}
	
	The results presented in this section are obtained with the network weights being trained on the minimum datasets ()\ref{table:min_dataset}). The evaluation of the network's performance after being trained on larger datasets (2000, 1000, 480, 320 and 240 grains ), followed to determine the minimum dataset size for each variation of the initial equation, is described in the Appendix A.

	\subsubsection{Angle distribution on the different test datasets}
	
	Initially, the test datasets produced with the three variations of the initial equation are analyzed. The focus of this analysis is on the distribution of the angles between the crack grain boundary edges and the horizontal axis. The study of the angle distribution on the test datasets provides important information on how extensive is the variation of these datasets compared to the initial dataset and enables a more efficient assessment of the predictions on the $J$-resistance values.
	
	\begin{figure}[!h]
		\centering
		\includegraphics[width = 0.95\linewidth]{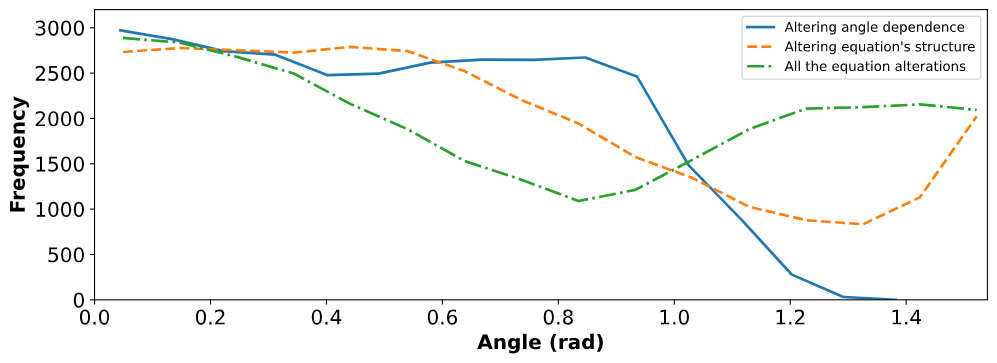}
		\caption{The grain boundary angle $\theta$ distribution in the test datasets, produced by the three variations of the Eq.~\eqref{eq:dj}.}
		\label{fig:angle_distrib_var}
	\end{figure}

	The plots in Fig.~\ref{fig:angle_distrib_var} show that the test datasets, produced by the three variations of the initial J-resistance equation, exhibit different distributions of the angles between the crack path edges and the horizontal axis. The alteration of the angle dependence on the Eq.~\eqref{eq:dj} does not significantly influence the angle distribution (solid line curve), when compared with the angle distribution in the test dataset produced by the initial equation (presented in Fig.~\ref{fig:angle_distrib_init}). On the other hand, the other two alterations of the initial equation produce datasets with substantial variation of the angle distributions (see dotted line curves in Fig.~\ref{fig:angle_distrib_var}). More specifically, studying the angle distribution of the dataset produced after the branching is removed from the initial equation, shows that once more the presence of the smaller angles is more pronounced in the crack path of the test dataset, but this time the test dataset also contains angles larger than 1.2 $rad$, which was not the case in the initial test dataset. These larger angles are present in the crack path because the equation that computes the crack growth resistance does not exhibit the branched structure and although the larger angles result into larger J-resistances, the energy difference is not as large as before. Even angles close to $90^{\circ}$ are present in the crack path with considerable probability. Even though this is incompatible with the experimental observations, since the crack very rarely follows an almost perpendicular trajectory when subjected to Mode-I loading, it serves the main objective of this study, which is to evaluate the performance of the network on datasets that deviate significantly from the initial dataset. 
	
	Finally, when both previous variations of the initial equation are implemented simultaneously, the crack path of the test dataset contains angles that cover the entire range from $0^{\circ}$ to $90^{\circ}$. The presence of large angles ($\theta > 1.2 rad$) in the crack path is even more pronounced, while the angle range that is less frequent in the dataset is between 0.6 rad to 1.0 rad. Evidently, the test datasets produced by the three variations of the initial $J$-resistance equation exhibit different characteristics and enable the evaluation of the performance of the algorithm on datasets with different local properties.     
	  
	\subsubsection{Evaluation of the meso-scale and micro-scale toughness prediction}
	
	After subjected to short training processes on the minimum datasets produced from the three variations of the initial energy equation, the performance of the network in predicting the meso-scale and micro-scale material toughness is assessed. 
	
	As before, the meso-scale toughness accuracy is evaluated by calculating the mean prediction accuracy of the $J$-resistance of the test dataset's crack segments.  Subsequently, the micro-scale toughness prediction accuracy is evaluated by calculating the accuracy of the more challenging predictions: the local energy increments ($dJ$) between two consecutive grain boundary junctions of the crack path.
	
	Thus, Table \ref{table:meso_micro_acc} presents the mean accuracy on the meso-scale and micro-scale toughness predictions on the test datasets produced by the three variations of the initial energy equation.  
	
	\begin{table}[!h]
		\centering
		\begin{tabular}[t]{ c |  c  c  c} 
			\toprule
			\toprule
			Equation variations         &  Meso-scale prediction accuracy	 &  Micro-scale prediction accuracy  \\ \toprule 
			Alter angle dependence      &           96.48\%                  &              83.72\%              \\ \midrule
			Alter equation's structure  &           94.84\%                  &              72.15\%              \\ \midrule
			Combine all the changes     &           95.56\%                  &              68.16\%              \\ \bottomrule  
			\bottomrule
		\end{tabular}
		\caption{The mean meso-scale and micro-scale toughness prediction accuracy values on the test datasets produced with three variations of the initial J-resistance equation are presented.}
		\label{table:meso_micro_acc} 
	\end{table}

	The evaluation of the meso-scale prediction accuracy shows that the neural network is capable of maintaining very high accuracy, after subjected on a limited training session with a dataset with size of 240 or 320 crack path grains. Considering that the meso-scale toughness of a crack path with this length can be obtained by \textbf{relatively simple experimental measurements}, this result is very important. This evaluation demonstrates that the graph neural network can extend its prediction performance on datasets that exhibit microstructural properties much different than the initial training dataset and the short additional training process requires a dataset that can be obtain from experimental measurements.
	
	The mean accuracy on the predictions of the local energy increments, which quantify the micro-scale toughness, is more sensitive to the extent of variation between the initial dataset and the dataset used to perform the additional training.  More specifically, in the case of the variation of the angle dependence, the prediction accuracy of the local energy increments on the test dataset is very high (83.72\%) and considering that the network is not trained to predict the local energy increments ($dJ$), this result reveals the great potential of the graph neural network in encoding the information of the graph components and capturing the inter-connections between its different components. 
	
	\begin{formal}
		The micro-scale toughness prediction accuracy on the other two datasets is lower, but the performance of the algorithm is still quite accurate. A closer look reveals the complexity of the task at hand and allows a better evaluation of these results.
	\end{formal}
	 
	 The structural variation of the Eq.~\eqref{eq:dj}, where the term $B_1 \theta - B_2$ is removed, computes very low values for the $J$-resistance (lower than $10^{-4} KJ/m^2$), when the angle between the grain boundary edge and the horizontal axis is close to $90^{\circ}$, due to the presence of the $cos(\theta)$ term. On the other hand, for small angles the same equation calculates $dJ$ values close to $1 KJ/m^2$. It is a very challenging task for the neural network to predict energy values that vary in such a large scale (from $1 KJ/m^2$ to $10^{-4} KJ/m^2$), especially when the network is not trained in predicting the local energy increments, but the total crack growth resistance of each crack path segment. During the training, the network learns how to predict the $J$-resistance values of the entire crack path segment, which is the sum of the local energy increments, and when summing the predictions on the local increments, these very low energy values have a minor influence on the final $J$-resistance prediction for the entire segment. Thus, for angles larger than 1.5 $rad$, the relative error of the predictions, compared to the ground truth, is very large; the $dJ$ value predicted by the network can be 10 or even 30 times larger than the ground truth. This results into negative prediction accuracy for angles larger than 1.5 $rad$ (or else larger than $86^{\circ}$). Nevertheless, this alteration of the Eq.~\eqref{eq:dj} is not compatible with the experimental data, since the propagation of the crack on a trajectory that follows a perpendicular path is very energy demanding and involves propagation modes different than mode I propagation, which is the focus of this work. The purpose of reformulating the equation in this manner that is incompatible to the experimental results is only to investigate the limitations of the graph neural network.

	Considering all these remarks and understanding that these negative accuracy values are obscuring the evaluation of the algorithm's performance, the predictions of the algorithm that correspond to angles higher than 1.5 $rad$ are removed from the computation of the mean prediction accuracy of the $dJ$ increments and from the plots presented in this section. As presented in the Table \ref{table:meso_micro_acc}, the computed mean micro-scale prediction accuracy of the algorithm is 72.15\% for the dataset produced by the structural variation of the initial equation and 68.16\% for the dataset that implements all the variations of the initial equation.

\begin{figure}[!h]
		\centering
		\includegraphics[width = 0.95\linewidth]{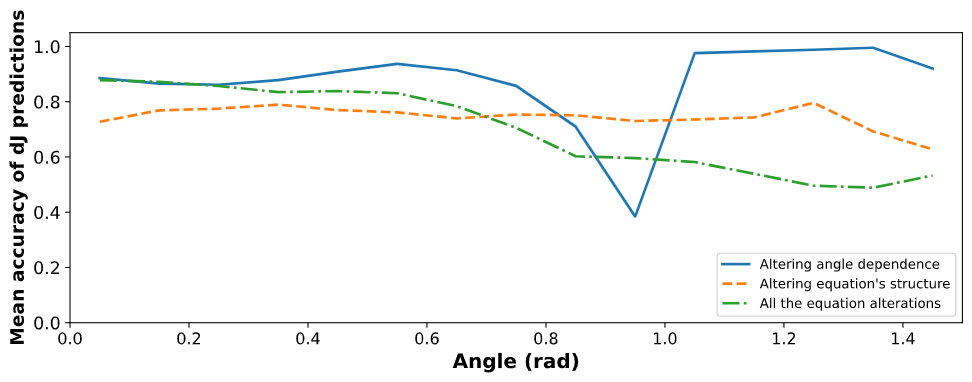} 
		\caption{The network's prediction accuracy on the local $dJ$ increments is plotted for different ranges of the angle between the grain boundary edges and the horizontal axis. These results are obtained after training the network on 3 different datasets, produced by variations of Eq.~\eqref{eq:dj}.}
		\label{fig:angle_acc}
	\end{figure}

	Additionally, the study of the relation of the prediction accuracy on the local energy increments ($dJ$) and the angle between the crack edge and the horizontal axis enables a better interpretation of the results. Fig.~\ref{fig:angle_acc} presents the mean accuracy in $dJ$ increments for different angle ranges of the test datasets, produced by the three equation variations.

	For the dataset produced after altering the angle dependence in Eq~\eqref{eq:dj}, the mean accuracy on the prediction of the local energy increments (presented with the solid line in Fig.~\ref{fig:angle_acc}) is high for most of the angle ranges (above 72\%), with the exception of the predictions made in grain boundary edges positioned in angles with the horizontal axis that belong to the range of  0.90 $rad$ to 1.0 $rad$. The center of this angle range is very close to the threshold value that defines the branching in the Eq.~\eqref{eq:dj} and apparently the computation of the local energy increments for grain boundary edges positioned in angles close to this threshold value is more challenging for the neural network. Similar issue was not observed when the network was trained on the initial dataset, as can be seen in Fig. \ref{fig:angle_acc_init}, since the weights of the network were trained on an extensive dataset.
	
	The prediction accuracy on the dataset produced with the structural variation of the initial equation is nearly independent on the angle between the grain edges and the horizontal axis and remains higher than 70\% in all the angle ranges. Finally, for the dataset that combines all the variations of the initial equation, the prediction accuracy is higher than 70\% for angles lower than 0.7 rad. As the angle between the grain boundary edge and the horizontal axis increases, the accuracy gradually decreases and reaches to 50\%. It is important that the prediction performance of the algorithm is high (above 70\%) for angles that are more relevant to the experimental observations, since the main objective is to create a computational framework that predicts the meso-scale and micro-scale toughness of different material systems, using data from experimental measurements.

	\begin{figure}[!h]
		\centering
		\begin{tabular}{ c c c}
			\toprule
			 & Altering angle dependence &  \\ \toprule
			\includegraphics[width = 0.29\linewidth]{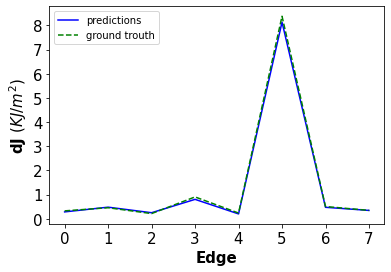} &
			\includegraphics[width = 0.29\linewidth]{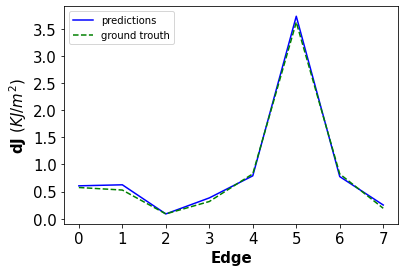} &
			\includegraphics[width = 0.29\linewidth]{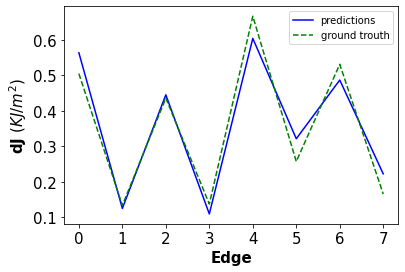}  \\ \toprule
			& Altering equation's structure & \\ \toprule
			\includegraphics[width = 0.29\linewidth]{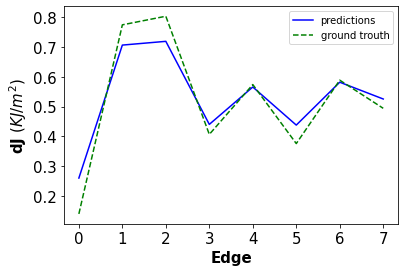} &
			\includegraphics[width = 0.29\linewidth]{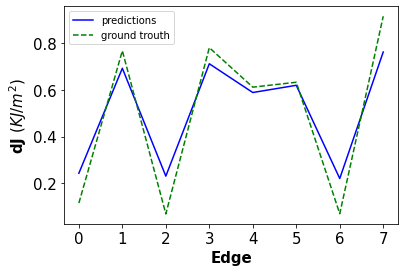} &
			\includegraphics[width = 0.29\linewidth]{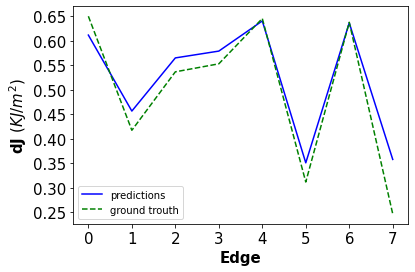}  \\ \toprule
			& Combining all changes & \\ \toprule
			\includegraphics[width = 0.29\linewidth]{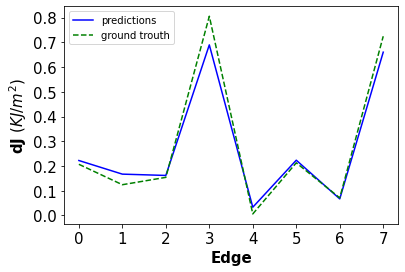} &
			\includegraphics[width = 0.29\linewidth]{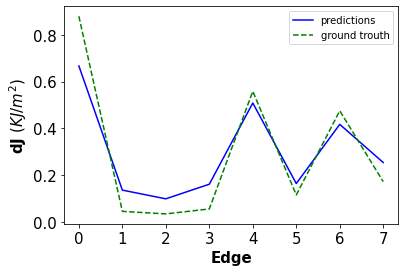} &
			\includegraphics[width = 0.29\linewidth]{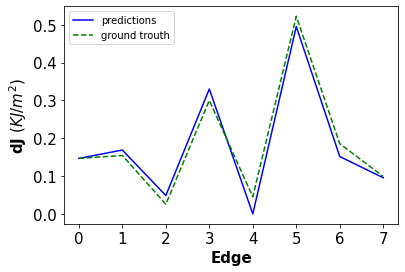}  \\ 
		\end{tabular}
		\caption{Ground truth and prediction values of the local $dJ$ increments between a grain boundary junction and the consecutive junction in the crack path. These results are obtained after training the network on a crack path with 240 or 320 grains, produced by the three variations of Eq.~\eqref{eq:dj}.}
		\label{fig:dj_var}
	\end{figure}

	Finally, in order to provide a better visualization of the micro-scale predictions, Fig.~\ref{fig:dj_var} presents some representative cases of the  $dJ$ predictions and their ground truth values for different segments of the test crack path. The plots in Fig.~\ref{fig:dj_var} show that the predictions of the neural network on the local energy increments do not deviate significantly from the ground truth values and the algorithm is capable of predicting the micro-scale variations of the crack growth resistance, even though the training process involves the minimization of the loss between the predicted $J$-resistance of the entire crack segment and their ground truth values (see Eq.~\eqref{eq:loss}). 
	
	The overall performance of the neural network in datasets that deviate significantly from the initial dataset is very good. The prediction accuracy of the meso-scale toughness is very high and in micro-scale level the accuracy is lower, but maintains high performance. These results show that the graph neural network, which is a key component of the computational framework introduced with this work (see Fig.~\ref{fig:flowchart}), exhibits great potential and the flexibility to be extended in performing accurate predictions on new datasets, after a short additional training on data that can be obtained from meso-scale experimental measurements.

	\subsection{Symbolic Regression}
	
	The functionality and efficiency of the Symbolic Regression part of the computational framework (see Fig. \ref{fig:flowchart}) is evaluated in this section. The Symbolic Regression algorithm enables the computation of an analytical mathematical formula that expresses better the relation between a set of variables. The objective is to define a mathematical expression in the form: $dJ = f(dr,\theta)$ that approximates the $dJ$ values obtained by the graph neural network predictions when the inputs are the length of the grain boundary edges and their angle. 
	
	To this end, the predictions of the network on the test dataset produced by the Eq.~\eqref{eq:dj_all}, which incorporates all the changes performed in the initial equation, and the corresponding geometrical properties of the test crack path are used to create a dataset ($[dJ,dr,\theta]$). This dataset is inserted to the Eureqa software \citep{eureqa}, which is using a Machine Learning algorithm to fit the input data ($dr,\theta$) to the output ($dJ$) and produces several different equations that are evaluated according to their complexity and accuracy in approximating the output values. 
	
	The equation, among the different options that Eureqa software computes, that combines a low level of complexity with high accuracy performance is selected and its accuracy in approximating the predicted by the graph neural network $dJ$ values is evaluated here. 
	
	This Symbolic Regression equation is presented in Eq.~\eqref{eq:eureqa}.   
	
	\begin{equation}
		dJ = \frac{C_1 + (C_2cos(\theta) - C_3\cdot \theta \cdot cos(C_4 - C_5 \cdot dr))}{C_6 + B^{(C_5 \cdot dr - C_7)}}
		\label{eq:eureqa}
	\end{equation}           
	
	The coefficients are defined as follows:
	
	\begin{table}[!h]
		\centering
		\begin{tabular}[t]{ c  c  c  c  c  c  c  c  c } 
			\toprule
			\toprule
			$B$   \hspace{3ex} &   $C_1$  \hspace{3ex} &   $C_2$ \hspace{3ex}  &  $C_3$  \hspace{3ex}  &  $C_4$ \hspace{3ex} &   $C_5$  \hspace{3ex} &   $C_6$  \hspace{3ex} &  $C_7$        \\ \toprule
			0.207078   \hspace{2.5ex} &   0.06626  \hspace{3ex} &   1.49 \hspace{3ex}  &  0.113  \hspace{3ex}   &  2.336 \hspace{3ex} &   152115 \hspace{3ex}  &   2.1368 \hspace{3ex}  &  2.167     \\ \toprule
			\bottomrule
		\end{tabular}
	\end{table} 
	
\begin{figure}[!h]
		\centering
		\begin{subfigure}[b]{0.5\textwidth}
			\centering
			\includegraphics[width = 0.9\linewidth]{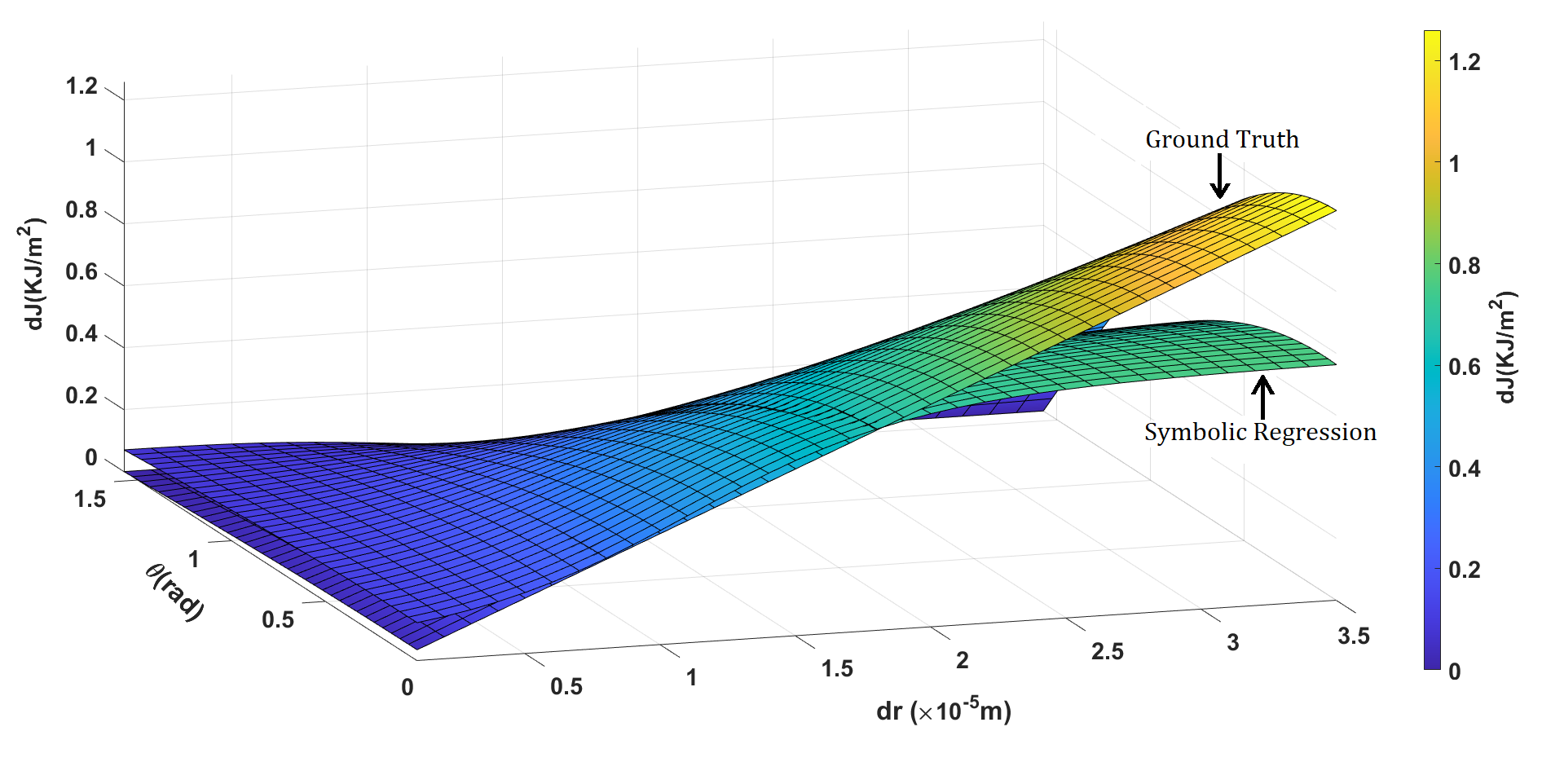}
			\caption{Symbolic regression results against ground truth.}
			\label{fig:eureqa_gt}
		\end{subfigure}%
		\begin{subfigure}[b]{0.5\textwidth}
			\centering
			\includegraphics[width = 0.9\linewidth]{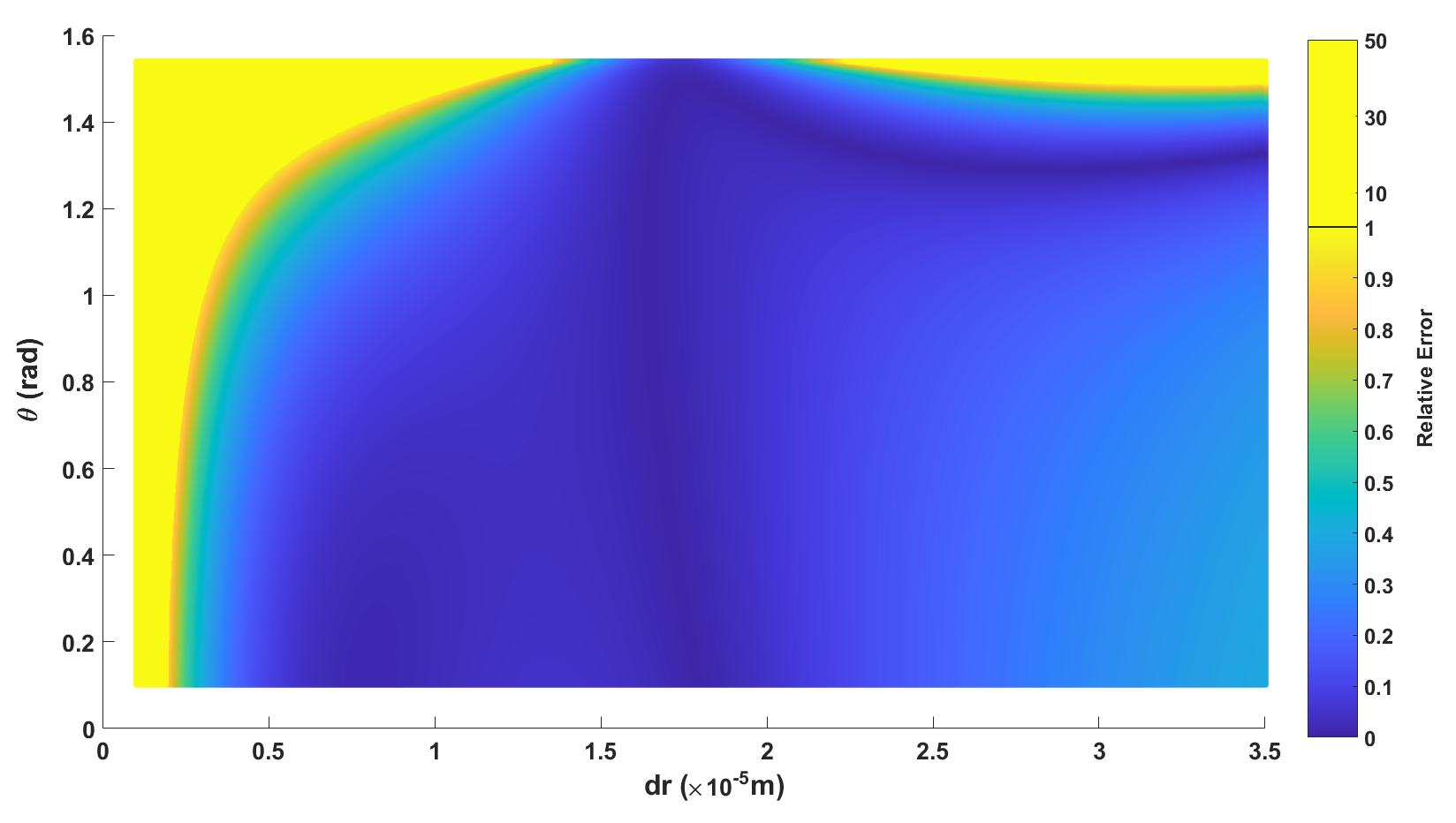}
			\caption{Relative error of symbolic regression results.}
			\label{fig:eureqa error}
		\end{subfigure}%
		\caption{Comparison between the $dJ$ values computed by the symbolic regression algorithm and the ground truth values obtained by Eq.~\eqref{eq:dj_all}.}
	\end{figure}

	The evaluation of the efficiency of the Symbolic Regression algorithm involves the comparison between the $dJ$ values computed by the Eq.~\eqref{eq:eureqa} to the ground truth values produced by the Eq.~\eqref{eq:dj_all}. To this end the two functions are plotted and the surface plots are presented in Fig. \ref{fig:eureqa_gt}. Additionally, the plot in Fig. \ref{fig:eureqa error} shows the relative error between the symbolic regression predictions and the ground truth.

	These plots show that the mathematical expression computed by the symbolic regression algorithm approximates the ground truth function efficiently in most of the computational domain. The highest relative error is observed in the very small numerical values, more specifically for $dJ_{gt} < 0.1 KJ/m^2$, the symbolic regression equation in some cases computes $dJ_{sym\_reg} \simeq 0.002KJ/m^2$ that results into relative error of 50. 
	
	Finally, it should be clarified that the error computed here can not be attributed solely to the symbolic regression results, but it also accounts for the error originated from the predictions of the neural network. The Eq.~\eqref{eq:dj_all}, which computes the ground truth values for this comparison, was implemented to generate the ground truth values in the test dataset, used as input for the graph neural network. On the other hand, the symbolic regression algorithm is using as input the $J$-resistance values that are predicted by the neural network on the test dataset. Thus, the relative error in the plots is the sum of the error in the graph neural network predictions and the error in the calculations of the symbolic regression equation.

	\section*{Summary}
	\label{conclusions_label}

	To conclude, a new computational framework is introduced with this publication, aiming to provide a valuable tool for extracting information of the material meso- and micro-scale mechanical properties. From experimental point of view this information is only accessible through extremely delicate experiments where the overall sampled volume and the experimental complexity limit the statistical assessment of the results and thus their validity.  Alternate approaches found in the literature, either rely on using finite elements analysis for approximating the solution to the inverse problem e.g.\citep{alabort2018grain} or aim at constructing reduced order models based on computational data that cannot be probed experimentally \citep{Hunter2018,mudunuru2019surrogate} on the relevant scale. 
	
	The efficiency of this computational tool in computing the micro-scale material toughness is very high and the expansion of its performance in material systems that exhibit different microstructural characteristics is enabled after a limited additional training. The size of the training dataset for this performance enhancement is investigated with different sizes of synthetic datasets, produced by a graph search algorithm, and it is proven that the amount of data required for an efficient training of the network can be obtained by simple experimental measurements.
	 
	Encouraged by the great potential of this computational tool, our forthcoming research work will be focused in addressing the following subjects:
	\begin{enumerate}
			\item The investigation of the performance of the algorithm on experimental data, and the evaluation of the associated limitations (e.g. measurement uncertainties, varying number of grains per segment etc.). The experimental validation of this computational tool is a very important next step and it will enable the study of specific phenomena, namely the degradation of grain boundaries due to chemical attack.
			\item The expansion of the performance of the framework in 3D data; the framework was only tested against 2D data, in order to keep the computational costs of generating the synthetic dataset low. In parallel, we aim to integrate effects such as the grain boundary character, plastic anisotropy, rate effects etc. to the functionality of the computational framework.   
	\end{enumerate} 
	
	By combining tools such as the one presented here, with existing experimental methodologies and other advanced analysis techniques, including quantitative fractography \citet{tsopanidis2020toward} and surrogate models \citet{alabort2018grain}, the authors believe that the fracture mechanics community, and specifically researchers in the community aiming to correlate materials microstructure with fracture experiments, are facing an exciting and promising future.

	\section*{Data availability}
	 The source code is available at \url{https://github.com/SteliosTsop/GNN_Material_Toughness_Predictor}.

	\section*{Acknowledgments}
	The financial support provided by the Pazy foundation young researchers award Grant \# 1176 is gratefully acknowledged.


	\bibliography{manuscript}
	\bibliographystyle{elsarticle-num-names}
	
	\clearpage


	\appendix

	\section{Investigation of the minimum dataset}

	\subsection{Altering the dependence on the grain boundary edge angle}

	Implementing this variation of the Eq.~\eqref{eq:dj}, four different training datasets are produced by the graph search algorithm. Each training dataset corresponds to a different crack path size, which is measured by the number of grains that the path contains. Similarly to the previous, each input crack path is divided into segments of 8 grains. These 4 training datasets, that contain 2000, 1000, 480 and 240 grains, are used as inputs for 4 independent training sessions that perform additional training on the network weights -- already trained on the extensive dataset produced by the original form of Eq.~\eqref{eq:dj}. 
	
	The mean accuracy on the prediction of J-resistance of the test dataset's crack segments is presented in the Table \ref{table:theta_total_accuracy}. 
	
	\begin{table}[!h]
		\centering
		\begin{tabular}[t]{ l |  c  c  c  c  } 
			\toprule
			\toprule
			\vtop{\hbox{\strut Number of crack path grains used }\hbox{\strut for the additional training.}}     &   2000   &   1000   &  480   &  240    \\ \toprule
			\vtop{\hbox{\strut Mean prediction accuracy of the}\hbox{\strut total $J$ value of the entire path.}} & 96.87\% & 96.53\% & 96.52\% & 96.48\% \\ \toprule
			\bottomrule
		\end{tabular}
		\caption{Mean prediction accuracy of the total J-resistance of the crack path segments of the test dataset, after the network is trained on crack path datasets produced by a variation of the Eq.~\eqref{eq:dj}, where the dependence of $dJ$ on the angle between the grain boundary edge and the horizontal axis is altered.}
		\label{table:theta_total_accuracy} 
	\end{table} 
	
	The network's prediction accuracy is not influenced by the size of the input crack path dataset, used for the additional training, and even a crack path that contains only 240 grains is sufficient to train the network efficiently. The $J$ accuracy exhibits very high values (96.48\%), even with the training on this small dataset.

	Similarly, the accuracy on the prediction of the $dJ$ increments for each crack edge, after training sessions with different input sizes of the training dataset, is computed and presented in the Table \ref{table:theta_local_accuracy}.
	
	\begin{table}[!h]
		\centering
		\begin{tabular}[t]{ l |  c  c  c  c  } 
			\toprule
			\toprule
			\vtop{\hbox{\strut Number of crack path grains used }\hbox{\strut for the additional training.}} &   2000   &   1000   &  480   &  240   \\ \toprule
			\vtop{\hbox{\strut Mean prediction accuracy on}\hbox{\strut the local $dJ$ increments.}} & 87.70\% & 83.75\% & 85.01\% & 83.72\%  \\ \toprule
			\bottomrule
		\end{tabular}
		\caption{Mean prediction accuracy of the local $dJ$ increments of each grain boundary edge in the crack path of the test dataset, after the network is trained on crack path datasets produced by a variation of the Eq.~\eqref{eq:dj}, where the dependence of $dJ$ on the angle between the grain boundary edge and the horizontal axis is altered.}
		\label{table:theta_local_accuracy} 
	\end{table} 
	
	The mean accuracy on the prediction of the local J-resistance increments for each crack path edge is very high, considering that the network is not trained on predicting the local energy increments. Even after the additional training with the smallest dataset, the accuracy on the prediction of the local $dJ$ increments is above 83\%.
	
\begin{figure}[!h]
		\centering
		\begin{tabular}{c c}
			\includegraphics[width = 0.48\linewidth]{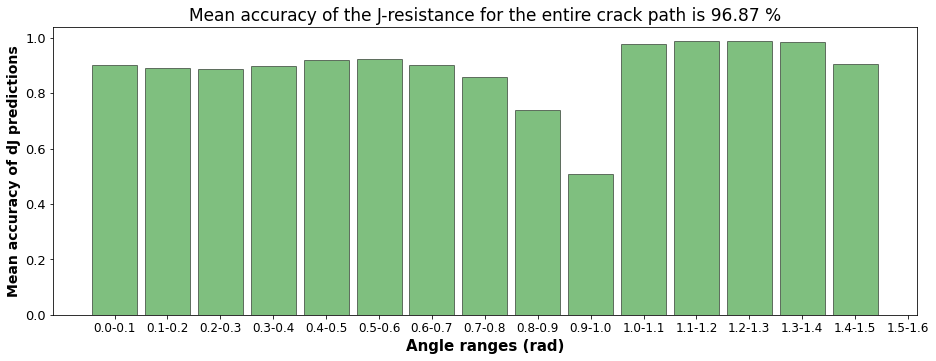} &
			\includegraphics[width = 0.48\linewidth]{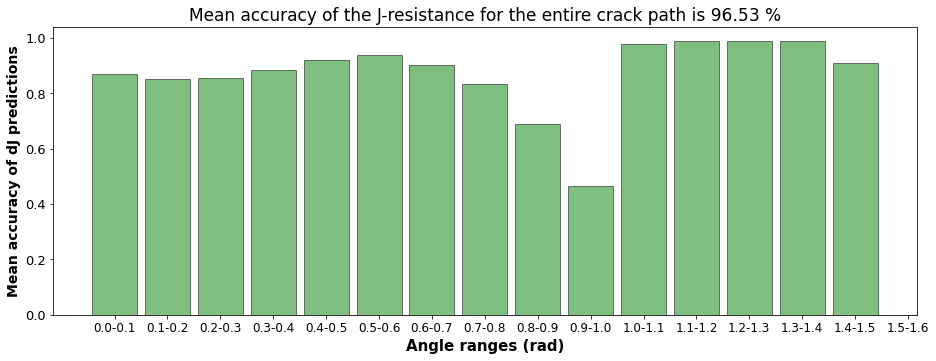} \\
			\scriptsize{a) Train with 2000 grains} & \scriptsize{b)Train with 1000 grains}\\
			\includegraphics[width = 0.48\linewidth]{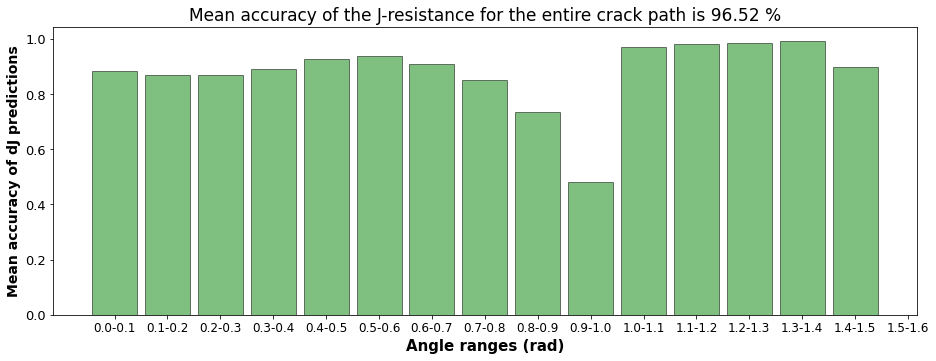} &
			\includegraphics[width = 0.48\linewidth]{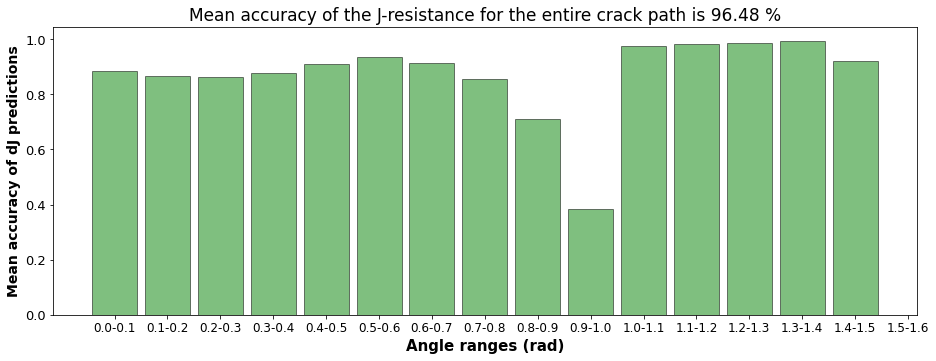} \\
			\scriptsize{c) Train with 480 grains} & \scriptsize{d)Train with 240 grains}\\
		\end{tabular}
		\caption{The network's prediction accuracy on the local $dJ$ increments is plotted for different ranges of the angle between the grain boundary edges and the horizontal axis. These results are obtained after training the network on 4 different datasets, produced by a variation of Eq.~\eqref{eq:dj} with respect to the dependence on the angle between the grain boundary edge and the horizontal axis.}
		\label{fig:theta_angle_acc}
	\end{figure}

	Finally, the relation of the prediction accuracy on the local energy increments and the angle between the crack edge and the horizontal axis is studied. The mean accuracy of the predictions for the $dJ$ increments for different consecutive ranges of the angle between the crack edge and the horizontal axis is computed and presented in the bar plots in Fig. \ref{fig:theta_angle_acc}.  
	
	\subsection{Altering the equation's structure}
	
	The Eq.~\eqref{eq:dj} is converted from a branch equation to a normal equation by removing the term $B_1 \theta - B_2$  and the performance of the network on a test dataset is evaluated. Similarly to the previous, five different training datasets are produced, implementing the altered Eq.~\eqref{eq:dj}, with 2000, 1000, 480, 320 and 240 grains. Consequently, the network is subjected to five independent training sessions, using as starting training point the pre-trained weights on the extensive initial dataset. 
	
	Following the previous evaluation process, the mean accuracy on the prediction of the J-resistance of the crack path segments is computed after each training session is concluded and the results are shown in Table \ref{table:nojump_total_accuracy}. The accuracy is very high and remains high even when the training is limited to 240 crack grains. 
	
	\begin{table}[!h]
		\centering
		\begin{tabular}[t]{ l |  c  c  c  c  c } 
			\toprule
			\toprule
			\vtop{\hbox{\strut Number of crack path grains used }\hbox{\strut for the additional training.}}     &   2000   &   1000   &  480   &  320  &   240    \\ \toprule
			\vtop{\hbox{\strut Mean prediction accuracy of the}\hbox{\strut total $J$ value of the entire path.}} & 95.86\% & 95.69\% & 95.54\% & 94.84\% & 93.43\%  \\ \toprule
			\bottomrule
		\end{tabular}
		\caption{Mean prediction accuracy of the total J-resistance of the entire crack path of the test dataset after the network is trained on crack path datasets produced by a variation of the Eq.~\eqref{eq:dj}, where the branch condition is removed.}
		\label{table:nojump_total_accuracy} 
	\end{table}

	Next, the accuracy on the prediction of the $dJ$ increments for each crack edge is evaluated. According to the values reported in Table \ref{table:nojump_local_accuracy}, the mean accuracy on the prediction of the local J-resistance increments for each crack path edge is influenced by the size of the training dataset and this is the reason of the increase of the lower limit size of the training dataset to 320 grains, instead of 240 grains that is used in the previous section. The prediction accuracy after training with a crack size of 320 grains is still good (72\%), while training with smaller datasets (240 grains) resulted in much lower accuracy (61.2\%).

	\begin{table}[!h]
		\centering
		\begin{tabular}[t]{ l |  c  c  c  c  c  } 
			\toprule
			\toprule
			\vtop{\hbox{\strut Number of crack path grains used }\hbox{\strut for the additional training.}} &   2000   &   1000   &  480   &  320 & 240  \\ \toprule
			\vtop{\hbox{\strut Mean prediction accuracy on}\hbox{\strut the local $dJ$ increments.}} & 82.97\% & 82.41\% & 72.11\% & 72.15\%  & 61.2\% \\ \toprule
			\bottomrule
		\end{tabular}
		\caption{Mean prediction accuracy of the local $dJ$ increments of each grain boundary edges of the crack path of the test dataset after the network is trained on crack path datasets produced by a variation of the Eq.~\eqref{eq:dj}, where the branch condition is removed.}
		\label{table:nojump_local_accuracy} 
	\end{table}
	
\begin{figure}[!h]
		\centering
		\begin{tabular}{c c}
			\includegraphics[width = 0.48\linewidth]{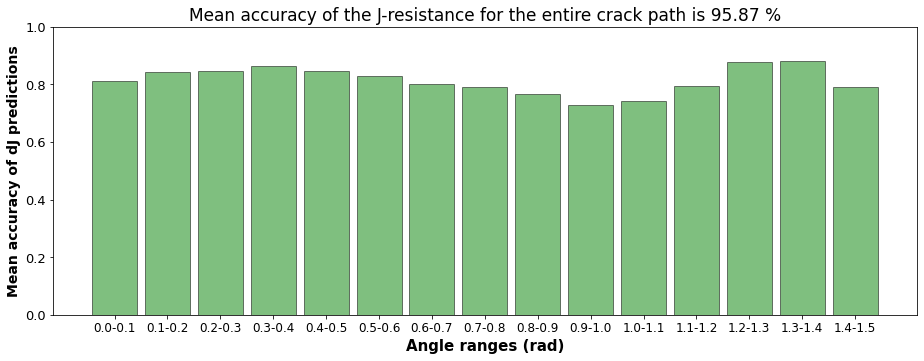} &
			\includegraphics[width = 0.48\linewidth]{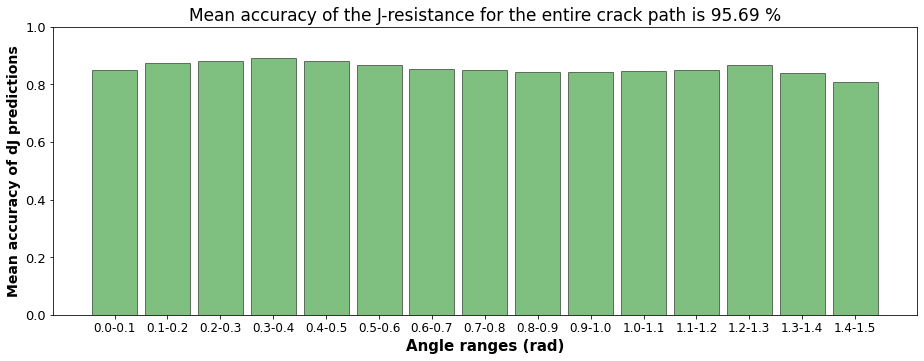} \\
			\scriptsize{a) Train with 2000 grains} & \scriptsize{b)Train with 1000 grains}\\
			\includegraphics[width = 0.48\linewidth]{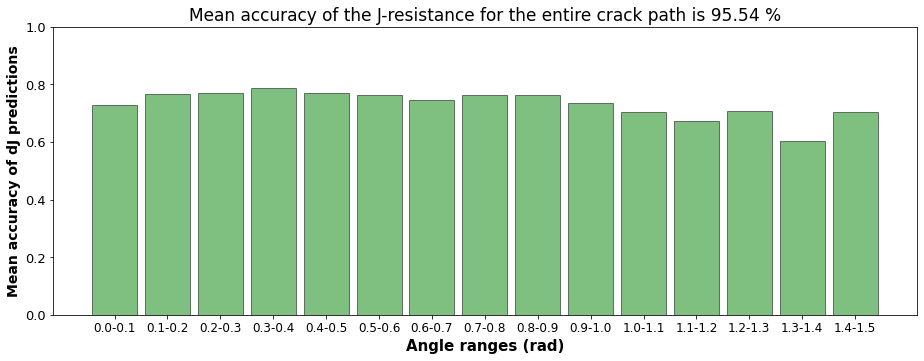} &
			\includegraphics[width = 0.48\linewidth]{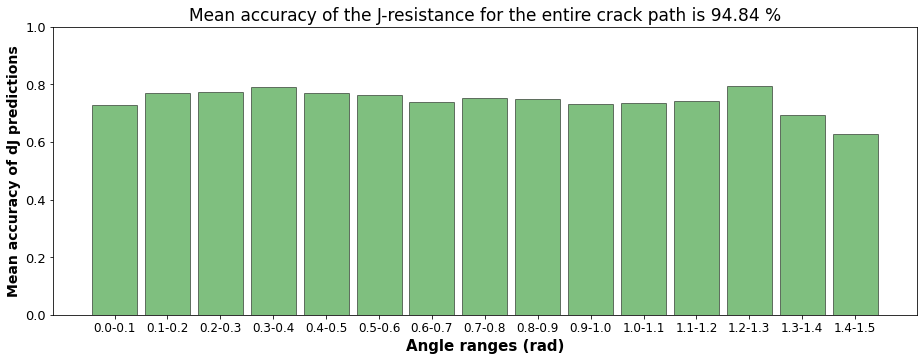} \\
			\scriptsize{c) Train with 480 grains} & \scriptsize{d)Train with 320 grains}\\
		\end{tabular}
		\caption{The network's prediction accuracy on the local $dJ$ increments is plotted for different ranges of the angle between the grain boundary edges and the horizontal axis. These results are obtained after training the network on 4 different datasets, produced by a reformulation of Eq.~\eqref{eq:dj} by removing the branching condition.}
		\label{fig:nojump_theta_angle_acc}
	\end{figure}

	Finally, the mean accuracy of the predictions for the $dJ$ increments for different consecutive ranges of the angle between the crack edge and the horizontal axis is computed and presented in the bar plots in Fig. \ref{fig:nojump_theta_angle_acc}.    
	
	These plots show that the accuracy remains high for all angle ranges and it is not significantly influenced by the size of the training dataset. 
	
	\subsection{Combining the previous changes}
	
	This section investigates the performance of the algorithm when all the previous changes are applied to the Eq.~\eqref{eq:dj}.
	
	Five training crack paths are produced with 2000, 1000, 480, 400 and 320 grains and different independent training cycles are performed with the same conditions as before. The network's performance after each training session is evaluated on a test dataset, generated with the same equation following the previous methodology.

	The mean accuracy on the prediction of the total J-resistance for the entire crack path segment is shown in Table \ref{table:all_total_accuracy} and it is very high, independently of the training dataset size. The general conclusion is that the network exhibits very high accuracy on the prediction of the total J-resistance of the input crack path segments, which is the task that it is trained to perform, and this conclusion is valid for every evaluation of the performance after training on different variations of the initial J-resistance equation. Even when all the previous changes are applied on the Eq.~\eqref{eq:dj}, the performance of the algorithm remains very high.
	
	\begin{table}[!h]
		\centering
		\begin{tabular}[t]{ l |  c  c  c  c  c  } 
			\toprule
			\toprule
			\vtop{\hbox{\strut Number of crack path grains used }\hbox{\strut for the additional training.}}     &   2000   &   1000   &  480   &  400  & 320  \\ \toprule
			\vtop{\hbox{\strut Mean prediction accuracy of the}\hbox{\strut total $J$ value of the entire path.}} & 96.52\% & 96.15\% & 95.71\% & 95.68\%  &  95.56\%  \\ \toprule
			\bottomrule
		\end{tabular}
		\caption{Mean prediction accuracy of the total J-resistance of the entire crack path of the test dataset after the network is trained on crack path datasets produced by a variation the Eq.~\eqref{eq:dj_all}.}
		\label{table:all_total_accuracy} 
	\end{table} 
	
	Subsequently, the performance of the network on the prediction of the local energy increments, which impose a greater challenge as the network is not trained for this operation, is evaluated. As before the negative prediction accuracy values that originate from the computation of the energy increments on grain boundary edges with angles close to $90^{\circ}$ ($\theta>1.5\; rad$) are removed from the computation of the mean accuracy. Table \ref{table:all_local_accuracy} presents the mean prediction accuracy on the local energy increments for the different training sessions of the network weights.   
	
	\begin{table}[!h]
		\centering
		\begin{tabular}[t]{ l |  c  c  c  c  c } 
			\toprule
			\toprule
			\vtop{\hbox{\strut Number of crack path grains used }\hbox{\strut for the additional training.}} &   2000   &   1000   &  480   & 400 & 320    \\ \toprule
			\vtop{\hbox{\strut Mean prediction accuracy on}\hbox{\strut the local $dJ$ increments.}}    & 79.24\%  & 77.31\% & 73.32\% & 72.57\% & 68.15\% \\ \toprule
			\bottomrule
		\end{tabular}
		\caption{Mean prediction accuracy of the local $dJ$ increments of each grain boundary edges of the crack path of the test dataset after the network is trained on crack path datasets, produced by a variation the Eq.~\eqref{eq:dj_all}.}
		\label{table:all_local_accuracy} 
	\end{table}

	Although the accuracy is lower compered to the accuracy reported in the previous sections, the efficiency of the network in predicting the local energy increments is still good, considering that the equation that computes the J-resistance has been substantially changed. The network after getting trained on a dataset with only 320 crack path grains computes the local energy increments with accuracy of 68\%, which is a good prediction accuracy for an operation that the network did not receive any training to perform.
	
\begin{figure}[!h]
		\centering
		\begin{tabular}{c c}
			\includegraphics[width = 0.48\linewidth]{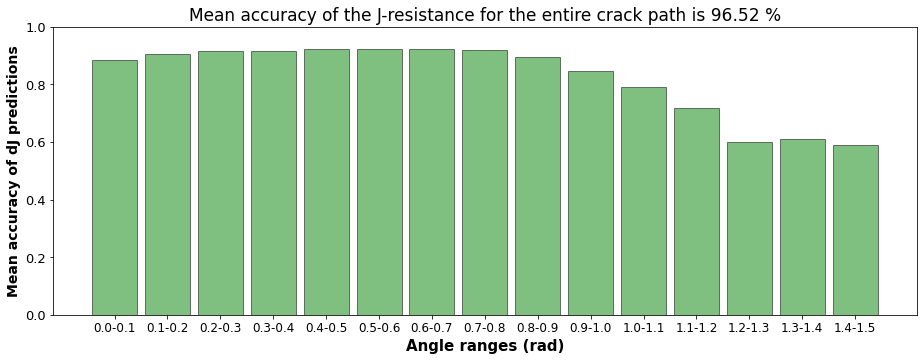} &
			\includegraphics[width = 0.48\linewidth]{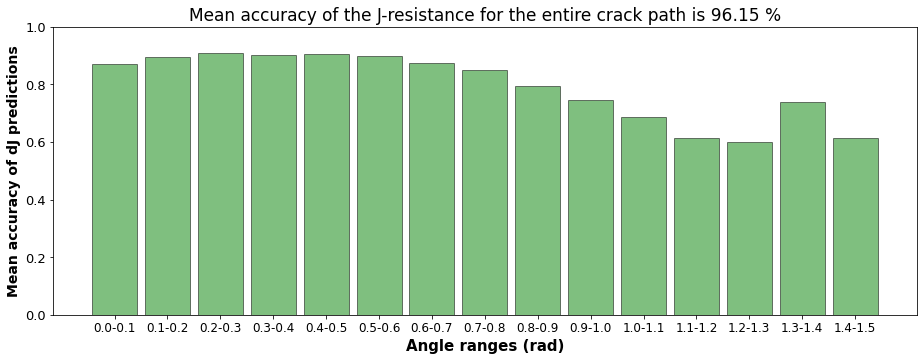} \\
			\scriptsize{a) Train with 2000 grains} & \scriptsize{b)Train with 1000 grains}\\
			\includegraphics[width = 0.48\linewidth]{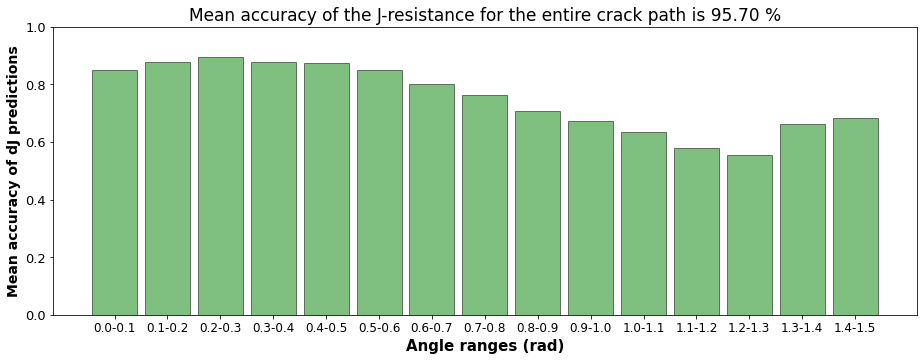} &
			\includegraphics[width = 0.48\linewidth]{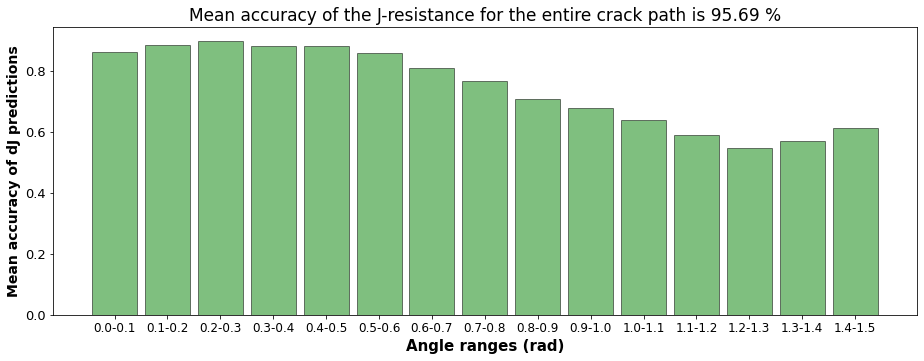} \\
			\scriptsize{c) Train with 480 grains} & \scriptsize{d)Train with 400 grains}\\
		\end{tabular}
		\caption{The network's prediction accuracy on the local $dJ$ increments is plotted for different ranges of the angle between the grain boundary edges and the horizontal axis. These results are obtained after training the network on 4 different datasets, produced by Eq.~\eqref{eq:dj_all}.}
		\label{fig:all_theta_angle_acc}
	\end{figure}

	Fig. \ref{fig:all_theta_angle_acc} presents the mean prediction accuracy for different ranges of angles between the grain boundary edges and the horizontal axis.

\end{document}